\documentstyle[pre,aps]{revtex}  
\begin{document}
\draft
\title{Local, Cluster, and Transitional Monte Carlo Dynamics\thanks{Talk
presented at the ``98 Workshop on Computational Material Science and
Biology'', 8--12, June 1998, Beijing.}}

\author{Jian-Sheng Wang}

\address{Department of Computational Science,
National University of Singapore,
Singapore 119260}

\date{15 July 1998}

\maketitle

\begin{abstract}

We review the local Monte Carlo dynamics and Swendsen-Wang
cluster algorithm.  We introduce and analyze a new Monte Carlo
dynamics known as transitional Monte Carlo.  The transitional Monte Carlo
algorithm samples energy probability distribution $P(E)$ with a
transition matrix obtained from single-spin-flip dynamics.  We analyze
the relaxation dynamics master equation,
$$    { d P(E, t) \over dt } = \sum_{E'} T(E,E') P(E',t), $$
associated with Ising model in $d$ dimensions.  In one dimension, we
obtain an exact solution.  We show in all dimensions in the continuum
limit the dynamics is governed by the partial differential equation
$$ {\partial P \over \partial t' } = 
            {\partial^2 P \over \partial x^2} + x {\partial P \over \partial x}
                + P. $$ 
where $x$ and $t'$ are rescaled energy deviation from the
equilibrium value and rescaled time, respectively.  This
equation is readily solved.  Thus, we have a complete understanding of
the dynamics.

\end{abstract}
\section{Monte Carlo Method}

Monte Carlo method \cite{Kalos,Binder} in the most basic application
is to perform numerical integration of very high dimensional integrals or
to compute averages of a given probability distribution.  In this
respect, the method generates a sequence of states $X_0$, $X_1$,
$X_2$, \dots, by a transition probability $T(X\to X') = P(X'|X)$.
This is also known as Markov chain Monte Carlo in the statistics
community.  The state $X$ can be the set of all the coordinates of the
particles in a fluid systems, or the values of spins at lattice points
for a classical spin system.  The new state $X'$ is generated
according to probability $P(X'|X)$ given that the previous state is
$X$.  $P(X'|X)$ is usually a simple distribution which can be sampled
directly.

We have a lot of freedom in choosing the transition probability $T(X
\to X')$.  If we want that the distribution of $X$ of the generated
sequence follows $P_{eq}(X)$, it is sufficient to require that
\begin{equation}
P_{eq}(X)\, T(X\to X') = P_{eq}(X')\, T(X' \to X).
\end{equation}
This is known as detailed balance condition.  Subject to some general
constraints of ergodicity which can be satisfied easily, the state $X$
has the unique equilibrium distribution in the large step limit.

Well-known application of Monte Carlo method is to generate Boltzmann
distribution $e^{-H/k_BT}/Z$ in statistical mechanics by Metropolis
algorithm, where the transition probability is chosen as
\begin{equation}
T(X \to X') =  W(X \to X') \min\left(1, {P_{eq}(X') \over
            P_{eq}(X) } \right), \quad X \neq X',
\end{equation}
where $W(X \to X') = W(X' \to X)$ is the probability of
proposing $X'$ as the next state given that the current state is $X$.
The next factor $\min(\cdots)$ is the probability that such a move is
accepted.
 
Averages of thermodynamic variables are computed according to
\begin{equation}
 \langle A \rangle = \sum_X A(X) P_{eq}(X) \approx { 1\over M} \sum_{i=1}^M A(X_i).
\end{equation}
The weighted sum (or integral) is replaced by an arithmetic average.

Monte Carlo method is intrinsically an approximate method.  Error in
Monte Carlo evaluation can be estimated from
\begin{equation}
   \delta A \approx { \sigma_A \over \sqrt{M/\tau} },
\end{equation}
where $\sigma_A^2$ is the variance of distribution of $A$, and $\tau$
is called correlation time.  The value $\tau \ge 1$ is more precisely
defined by so-called integrated correlation time.  But in many cases
it is equivalent to the exponential correlation time defined by the
equation:
\begin{equation}
  \langle A(t_0) A(t_0 + t) \rangle - 
            \langle A(t_0) \rangle^2 \propto e^{-t/\tau}.
\end{equation}
The exponential correlation time $\tau$ also characterizes the speed
at which arbitrary initial probability distribution $P(X)$ converges
to $P_{eq}(X)$:
\begin{equation}
   P(X,t) \approx P_{eq}(X) + P_1(X) e^{-t/\tau} + \cdots.
\end{equation}

\section{Some Monte Carlo Dynamics}
The above theory is quite general.  We give some more concrete
examples and points out some interesting Monte Carlo dynamics.

\subsection{Single-spin-flip Glauber dynamics}
We consider a simple classical spin model, the Ising model, as an
example of local Monte Carlo dynamics.  The model is defined by the
energy function
\begin{equation}
      H(\sigma) = - J \sum_{\langle i,j\rangle}\sigma_i \sigma_j, \qquad 
  \sigma_i = \pm 1.
\end{equation}
The spins take only two possible values and live on the sites of a
lattice, for example on a square lattice.  The total energy is a sum
of interactions between nearest-neighbor sites.  A Monte Carlo move
consists of picking a site at random, and flipping the spin with
probability \cite{Glauber}
\begin{equation}
 w = { 1\over 2} \left[ 1 - \sigma_0 \tanh\left( {J\over k_B T} \sum_i 
\sigma_i \right) \right], \label{weq}
\end{equation}
where $\sigma_0$ is the spin value before the flip, and $\sum_i \sigma_i$
is the sum of spins of the nearest neighbors.  Another popular choice is
the Metropolis rate $\min\bigl(1, \exp(-\Delta H/k_B T) \bigr)$ where
$\Delta H$ is the energy increase due to flip.  In a computer
implementation, the value $w$ is compared with a uniformly
distributed random number $r$ between 0 and 1. If $r < w$ the flip is
performed; otherwise, the old configuration is retained and is also
counted as one Monte Carlo move.  One Monte Carlo step is a unit
(Monte Carlo) time, and is usually defined as $N$ Monte Carlo moves,
where $N$ is the number of spins of the system.

The local Monte Carlo dynamics has some common features: (1) the
algorithm is extremely general.  It can be applied to any classical
model.  (2) Each move involves $O(1)$ operation and $O(1)$ degrees of
freedom.  (3) At the second-order phase transition critical point, the
dynamics becomes very slow.  This is known as dynamical critical slowing
down, characterized by the fact that
\begin{equation}
   \tau \propto \xi^z, \qquad z \approx 2.
\end{equation}
where $\xi \propto | T - T_c |^{-\nu}$ is called correlation length
which will diverge at the critical point.  On finite system, $\xi$ is
replaced by the system linear size $L$.  The value $z$ is the
dynamical critical exponent.  Its value is slightly greater than
2 for a large class of
models with order-parameter nonconserving dynamics, such as the
single-spin-flip dynamics discussed above \cite{Wang-Gan}.

This last feature hampers the effective use of local Monte Carlo
algorithms.  It is the nonlocal algorithms that come to rescue.

\subsection{Nonlocal dynamics --- cluster algorithms}

Swendsen-Wang algorithm
\cite{Swendsen-Wang,Swendsen-Wang-Ferrenberg} is one of the
first nonlocal Monte Carlo algorithms which have very different
dynamical characteristics.  The algorithm uses a mapping from Ising
model to a type of percolation model.  Each Monte Carlo step consists
of putting a bond with probability
\begin{equation}
    p(\sigma_i, \sigma_j) = 1 - 
        \exp\bigl( -J(\sigma_i \sigma_j + 1)/k_B T\bigr)
\end{equation}
between each pair of the nearest neighbors.  by ignoring the spins and
looking only at the bonds, we obtain a percolation configurations of
bonds \cite{Stauffer}.  A new spin configuration is obtained by
assigning to each cluster, including isolated sites, a random sign
$+1$ or $-1$ with equal probability.

In the Swendsen-Wang algorithm, we generated many clusters and then
flipped these clusters.  Wolff algorithm \cite{Wolff} is a variation on
the way clusters are flipped.  One picks a site at random, and then
generates one single cluster by growing a cluster from the seed.  The
neighbors of the seed site will also belong to the cluster if the
spins are in parallel and a random number is less than $p = 1 -
e^{-2J/k_B T}\!$.  That is, the neighboring site will be in the same
cluster as the seed site with probability $p$ if the spins have the
same sign.  If the spins are different, neighboring site will never
belong to the cluster.  Neighbors of each new site in the cluster are
tested for membership.  This testing of membership is performed on
pair of sites (forming a nearest neighbor bond) not more than once.
The recursive process will eventually terminate.  The spins in the
cluster are turned over with probability 1. 

The following is a fairly elegant way of implementing the Wolff
algorithm in C.  The function is the core part which performs a Wolff
single cluster flip.  This function is recursive.  The array for spins
{\tt s[$\;$]}, percolation probability {\tt P}, and coordination
number {\tt Z} (the number of neighbors) are passed globally.  The
first argument {\tt i} of the {\tt flip} function is the site to be
flipped, the second argument {\tt s0} is the spin of the cluster
before flipping. The function {\tt neighbor} returns an array of
neighbors of the current site.  The function {\tt drand48()} returns
a random number unformly distributed between 0 and 1.

{
\font\ninerm=cmr9
\font\ninett=cmtt9
\def\l{\hbox to \hsize}
\def\m#1{\hfill{\ninerm #1}\hskip 1cm}
\def\undertext#1{$\underline{\smash{\hbox{#1}}}$}
\def\uncatcodespecials{\def\do##1{\catcode`##1=12 }\dospecials}
{\obeyspaces\global\let =\ } 
{\catcode`\`=\active \gdef`{\relax\lq}}
\def\v{\begingroup\hskip1cm\tt\uncatcodespecials
  \obeyspaces\doverbatt}
\newcount\balance
{\catcode`<=1 \catcode`>=2 \catcode`\{=12 \catcode`\}=12
  \gdef\doverbatt`<\balance=1\verbatimloop>
  \gdef\verbatimloop#1<\def\next<#1\verbatimloop>%
    \if#1`\advance\balance by-1
     \ifnum\balance=0\let\next=\endgroup\fi\fi\next>>
\vbox{
\smallbreak
\l{\v`void flip(int i, int s0)`\m{}}
\l{\v`{`\m{}}
\l{\v`   int j, nn[Z];`\m{}}
\smallskip
\l{\v`   s[i] = - s0;`\m{flip the spin immediately}}
\l{\v`   neighbor(i, nn);`\m{find nearest neighbor of {\ninett i}}}
\l{\v`   for(j = 0; j < Z; ++j)`\m{flip the neighbor if}}
\l{\v`      if(s0 == s[nn[j]] && drand48() < P)`\m{spins are equal and}}
\l{\v`         flip(nn[j], s0);`\m{random number is smaller than {\ninett p}}}
\l{\v`}`\m{}}
\smallbreak
}
}

Some of the salient features of cluster algorithms: 
(1) algorithm is applicable to models
containing Ising symmetry. (2) Computational complexity is still of
$O(1)$ per spin per Monte Carlo step.  (3) Much reduced critical
slowing down, i.e., $\tau \propto \xi^{z_{sw}}$.  The dynamical
critical exponent $z_{sw}$ is 0, 0.3, 0.5, 1, in dimensions 1, 2, 3,
and greater or equal to 4, respectively.  In addition, Li and Sokal
\cite{Li-Sokal} showed that $\tau \ge a c$ for some constant $a$, 
and $c$ is the specific heat. For some of the lastest developments
on nonlocal algorithms, see \cite{Evertz,Krauth,Machta,Beard}.

\section{Transitional Monte Carlo Dynamics}

The transitional Monte Carlo dynamics \cite{Wang-Tay-Swendsen}
is a new dynamics with the following interesting features:
(1) It is a constrained random walk in energy space.  (2)
The transitional rates are derived from single-spin-flip dynamics.
(3) It has a fast dynamics, $\tau \propto c$, and (4) it suggests
a different histogram reweighting method.  

\subsection{What is transitional Monte Carlo}

The transitional Monte Carlo is related to the single-spin-flip
dynamics in the following sense but it has a totally different
dynamics.  A single-spin-flip Glauber dynamics of the Ising model is
described by
\begin{eqnarray}
{\partial P(\sigma, t) \over \partial t} =  &&
\sum_{\{\sigma'\}} \Gamma(\sigma,\sigma') P(\sigma', t) \nonumber \\
 = && \sum_{i=1}^N \Bigl[ -w_i(\sigma_i) + w_i(-\sigma_i)F_i\Bigr] P(\sigma,t),
\label{Meq}
\end{eqnarray}
where $N$ is the total number of spins, and $w$ is given by Eq.~(\ref{weq}),
and $F_i$ is a flip operator such that $F_i
P(\ldots,\sigma_i, \ldots) = P(\ldots,-\sigma_i, \ldots)$.
Transitional Monte Carlo dynamics is {\sl defined} by
\begin{equation}
 { d P(E,t) \over dt }  = \sum_{E'} T(E,E') P(E',t), \label{Teq}
\end{equation} 
where $P(E,t)$ is the probability of having energy $E$ at time $t$, and 
\begin{equation}
  T(E,E') = {1 \over n(E') } \sum_{H(\sigma) = E}\sum_{H(\sigma') = E'}
\!\!\! \Gamma(\sigma, \sigma'),
\end{equation} 
where $n(E)$ is the degeneracy of the states.  We can not derive
Eq.~(\ref{Teq}) from Eq.~(\ref{Meq}) in general, the ``derivation'' is
valid only at equilibrium when $P(E) = \sum_{H(\sigma) = E} P(\sigma)
= n(E) \exp(-E/k_BT)$.

The transition matrix $T(E,E')$ has some general properties: (1) The
matrix is banded alone the diagonal.  (2) The column sum is zero,
$\sum_E T(E,E') = 0$, due to the conservation of total probability.
(3) $\sum_{E'} T(E,E') P_{eq}(E') = 0$, due to existence of
equilibrium distribution.  (4) The transition rate satisfies detailed
balance condition, $T(E',E) P_{eq}(E) = T(E,E') P_{eq}(E')$.  The
eigenvalues of $T(E,E')$ are real and $\lambda_k \leq 0$.  The
significance of the eigenspectrum $\lambda_k$ is that the general
solution of the master equation can be written as $P(E,t) = \sum_k
a_k(E) \exp(\lambda_k t)$.

\subsection{Computer realization of transitional Monte Carlo and 
reweighting techniques}

The transitional Monte Carlo dynamics can be implemented on computer
in at least two different ways, we'll call them algorithm A and B.

\medskip
{\noindent\bf Algorithm A}

\begin{enumerate}
  \item Do sufficient number of constant energy (microcanonical) 
Monte Carlo steps, so that the final configuration is totally uncorrelated
with the initial configuration.  This step is equivalent to pick a state
$\sigma$ at random from all states with energy $E$.

   \item Do one canonical Monte Carlo move by picking a site at random.
\end{enumerate}
Clearly, this algorithm is not very efficient computationally, due to
step 1.  However, it will be helpful in understanding the dynamics.

\medskip
{\noindent\bf Algorithm B}

\begin{itemize}
\item A direct implementation of the master Eq.~(\ref{Teq}), i.e., a random
walk in energy with a transition rate $T(E,E')$. 
\end{itemize}

Other possibility is to solve the equation on computer.  Then in this method
and algorithm B, we need to know $T(E,E')$ explicitly, this can be
done numerically by Monte Carlo sampling, from
\begin{equation}
  T(E+\Delta E,E) = w(\Delta E) 
\bigl\langle N(\sigma, \Delta E) \bigr\rangle_{H(\sigma)=E}
\end{equation}
and $w(\Delta E) = {1\over 2} \bigl( 1 - \tanh(\Delta
E/(2k_BT)\bigr)$.  $N(\sigma, \Delta E)$ is the number of cases that
energy is changed by $\Delta E$ from $E$ for the $N$ possible
single-spin flips.

Note that computation of $\bigl\langle N(\sigma, \Delta E)
\bigr\rangle_{H(\sigma)=E}$ can be done with any sampling technique
which ensures equal probability for equal energy.  It is a kind of
``combinatorial'' number independent of the spin flip rates and in
particular, independent of the temperature.  Thus the transition
matrix can be formed with any temperature.  The equilibrium
distribution and thus the density of states $n(E) = P_{eq}(E)
\exp(E/k_BT)$ is obtained by solving
\begin{equation}
   \sum_{E'} T(E,E') P_{eq}(E') = 0.
\end{equation}
The above scheme is similar in spirit to the histogram method of
Ferrenberg and Swendsen \cite{Ferrenberg-Swendsen}, and the method has
a close connection with, but different from the broad histogram of
Oliveira et al \cite{Oliveira}.

\subsection{Exact results in transitional Monte Carlo dynamics}

We have more or less a complete understanding of the transitional
Monte Carlo dynamics through exact results in limiting cases.
The transition matrix $T(E,E')$ can be computed
exactly in one-dimensional chain of length $L$ (with periodic 
boundary condition), by some combinatorial consideration, as 
\begin{eqnarray}
    T_{k, k+1} = && { (k+1)(2k+1) \over L - 1} (1+\gamma), \\
    T_{k+1,k} = && { (L-2k)(L-2k-1) \over 2 (L-1) } (1-\gamma),
\end{eqnarray}
where $\gamma = \tanh(2J/k_BT)$.
The diagonal term is computed from the relation
\begin{equation}
    T_{k-1,k} + T_{k,k} + T_{k+1,k} = 0,
\end{equation}
and the rest of the elements $T_{k,k'} = 0$ if $|k-k'| > 1$.  The
integer $k=0, 1, 2, \ldots, \lfloor L/2 \rfloor$ is related to energy
by $E= -LJ + 4k$.  While the eigen spectrum at temperature $T=0$ can
be computed exactly as $\lambda_k = - 2(k+1)(2k+1)/(L-1)$, the
eigenvalues at $T>0$ is obtained only numerically.  The most important
feature is that $\tau \propto L$, given an unusual dynamical critical
exponent of $z=1$.

The dynamics in any dimensions in the large size limit obeys a linear
Fokker-Planck equation:
\begin{equation}
   {\partial P(x,t') \over \partial t'} = 
{ \partial \over \partial x } \left( {\partial P(x,t') \over \partial x }
+ x P(x,t') \right), \label{diffusionEq}
\end{equation}
where $t'$ and $x$ are properly scaled time and energy.   
\begin{equation} 
  x = { E - u_0 N \over ( N c' )^{1/2} },  \quad u_0 N = \bar E,
\end{equation}
and $t' = b t$ with
\begin{equation}
  b = \lim_{N\to\infty} {1\over 2 c' N} \sum_{E'} T(\bar E,E') (E'-\bar E)^2,
\label{beq}
\end{equation}
where $u_0$ is the average energy per spin and $c'=k_BT^2c$ is the
reduced specific heat per spin.  The major consequence of this result
is that the relaxation times are $\tau_n = a c'/n$, $n=1,2,3,\cdots$,
with some constant $a$.

We can cast the equation in the form of a continuity equation,
\begin{equation}
  { \partial P \over \partial t } + {\partial j \over \partial x} = 0,
\quad \hbox{with} \quad j = - {\partial P\over \partial x} - x P.
\end{equation}
There are two competing effects in the current; $-\partial P/\partial
x$ is the usual diffusion, while $-xP$ is a counter drift.  $j=0$
produces the equilibrium distribution $P_{eq}(x) \propto
\exp(-x^2/2)$.

With a change of variable $P(x,t') = \exp(-x^2/4) \Phi(x,t')$, the
Eq.~(\ref{diffusionEq}) is transformed into a one-dimensional quantum
harmonic oscillator equation, with a general solution of the form
\begin{equation}
   \sum_{n=0}^\infty c_n \exp(-nbt - x^2/2) H_n(x/\sqrt{2}),
\end{equation}
where $H_n$ is Hermite polynomials.

We sketch the derivation of the continuum limit equation, which is
known as $\Omega$ expansion \cite{VanKampen}.  Starting from the
master equation, Eq.~(\ref{Teq}), we introduce the new variable $x$
which describes the scaled deviations from equilibrium,
\begin{equation}
   E = \bar E + \delta E = u_0 N + x (Nc')^{1/2}.
\end{equation}
The function $P(E,t)$ is written in terms of $x$, and $P(E',t) \to
P(x+\delta x, t) = P(x,t) + {\partial P \over \partial x} \delta x +
{1 \over 2} {\partial^2 P \over \partial x^2 } \delta x^2 + \cdots$.
For the matrix $T(E,E')$, we assume that the changes along the
diagonal are smooth and can be expanded, but across diagonals the
changes are still discrete.  For $T(E,E')$ we also expand around
$x=0$; we can show that such an expansion is also an expansion in
power of $N^{-1/2}$.  Expanding all the relevant terms in powers of
$N^{-1/2}$, the leading terms of order $N^0$ give the desired
equation. The rest of the correction terms go to zero in the large
size limit $N \to \infty$.  We used some of the general properties of
$T(E,E')$ to simplify the equation.

\subsection{Simple pictures of the dynamics}

The exact results can be interpreted with intuitive pictures.  First,
we consider the result of $\tau \propto L$ in one dimension as $T \to
0$.  At sufficiently low temperatures with a correlation length $\xi$
comparable with the system size $L$, only the ground state (all spins
up or down) and the first excited states (with a kink pair) are
important.  Let's consider the time scale for $E_0 \to E_1$, a spin
with opposite sign is created with probability $\exp(-4K)$ from
Boltzmann weight, where $K=J/(k_BT)$, in each of the canonical move.
Thus
\begin{equation}
   \tau \propto { \exp(4K) \over L } \propto {\xi^2 \over L} \propto L.
\end{equation}
where $K$ is chosen such that there is about one kink pair, so
that $\xi \sim \exp(2K) \sim L$.

The reverse process, $E_1 \to E_0$ has also the same time scale.  In
this case, it is advantageous to use the equivalence of Algorithm A and
B.  Given that the system is in the state of kink pair (a string of up
spins followed by a string of down spins, with periodic boundary
condition), the first step of Algorithm A randomizes the locations of
the kinks.  The probability that two kinks are the nearest neighbors
is $1/L$; the probability that this pair is chosen and destroyed by a
spin flip is $1/L$.  Thus, the transitional Monte Carlo moves needed
to destroy a kink pair are $1/((1/L)(1/L)) = L^2$.  The time in terms
of Monte Carlo step is then $\tau \propto L^2/L = L$.

Similarly, the result of $\tau \propto c$ can be obtained by the
following argument.  The transitional Monte Carlo is a random walk
constrained in the range $\delta E$, due to the gaussian distribution
nature of the equilibrium distribution $P_{eq}(E)$.  The width of this
distribution is related to the specific heat by $\delta E^2 = c N
k_BT^2$.  Each walk changes $E$ by $O(1)$. To change $E$ by $\delta E$,
we need 
$\delta E^2$ moves, invoking the theory on random walks.
In units of transitional Monte Carlo steps,
\begin{equation}
  \tau \approx a { \delta E^2 \over N} \propto  c.
\end{equation}

\section*{Acknowledgements}

Part of the work presented in this talk is in collaboration with
Tay Tien Kiat and Robert~H.~Swendsen.  This work is supported in part by
a NUS Faculty Research Grant PR950601.

\end{document}